\documentclass[preprint,showpacs,preprintnumbers,amsmath,amssymb]{revtex4}

\usepackage{graphicx}
\usepackage{dcolumn}
\usepackage{bm}

\begin{document}

\preprint{APS/}

\title{Effect of chromatic dispersion induced chirp on the temporal coherence property of individual beam from spontaneous four wave mixing}

\author{Xiaoxin Ma, Xiaoying Li }
\email[]{xiaoyingli@tju.edu.cn}
\author{Liang Cui, Xueshi Guo, and Lei Yang}

\address{ College of Precision Instrument and
Opto-electronics Engineering, Tianjin University, \\Key Laboratory of Optoelectronics Information Technology, Ministry of Education, Tianjin, 300072,
P. R. China}

\date{\today}

\begin{abstract}
Temporal coherence of individual signal or idler beam, determined by the spectral correlation property of photon pairs, is important for realizing quantum interference among independent sources. To understand the effect of chirp on the temporal coherence property, two series of experiments are investigated by introducing different amount of chirp into either the pulsed pump or individual signal (idler) beam. In the first one, based on spontaneous four wave mixing in a piece of optical fiber, the intensity correlation function of the filtered individual signal beam, which characterizes the degree of temporal coherence, is measured as a function of the chirp of pump. The results demonstrate that the chirp of pump pulses decreases the degree of temporal coherence. In the second one, a Hong-Ou-Mandel type two-photon interference experiment with the signal beams generated in two different fibers is carried out. The results illustrate that the chirp of individual beam does not change the temporal coherence degree, but affect the temporal mode matching. To achieve high visibility, apart from improving the coherence degree by minimizing the chirp of pump, mode matching should be optimized by managing the chirps of individual beams.

\end{abstract}

\pacs{42.50.Dv, 42.65.Lm, 03.67.Hk}
\maketitle

\section{Introduction}
Quantum correlated signal and idler photon pairs from parametric processes have been a crucial resource for quantum metrology and quantum information processing. To successfully fulfill a specific task, one needs to consider the spectral (temporal) properties. Particularly, for the tasks involving quantum interference among independent sources, such as quantum teleportation and linear optical quantum
computing~\cite{Bou97,KLM}, photon pairs in a spectral factorable state are highly desirable~\cite{Uren05,Palmett07,Mosley08}. In this case,
individual signal or idler beam (photons) in single mode thermal state is said to have high temporal coherence, which is the key to form quantum interference with high visibility~\cite{ou99}.


Generally speaking, the coherence degree of individual beam, determined by the correlation property of photon pairs, is related to its spatial and temporal modes. Because it has been proved that the individual beam exhibits thermal fluctuation~\cite{Yurke87}, and there is a relation between the measured bunching coefficient and the mode number of a thermal field ~\cite{Wasilewski08}, we characterize the degree of coherence by measuring the intensity correlation function $g^{(2)}$~\cite{ou99}. Moreover, photon pairs generated from spontaneous parametric emissions (SPE), including spontaneous parametric down conversion (SPDC) and spontaneous four wave mixing (SFWM) in $\chi ^{(2)}$ and $\chi ^{(3)}$ nonlinear media, respectively, usually exhibit significant spectral and spatial correlations. Considering the spatial
correlations can be minimized by using guided-wave configurations, here we focus on the spectral correlation determined temporal coherence property.

When $g^{(2)}$ of individual signal or idler field is measured by using single photon detectors (SPDs) with response time much longer than the coherence time of the thermal field, $g^{(2)}$ increases with the decrease of the mode number contained in the field. For instance,
for photon pairs in spectral factorable state, the measured photon statistics of the single mode individual beam is $P(n)=\frac{{\overline{n}^n}}{{(1+\overline{n})^{n+1}}}$ (Bose-Einstein distribution), where $\overline{n}$ and $P(n)$ denote the average photon number and probability of detecting n-photon, respectively. In contrast, for the photon pairs with perfect spectral correlation, individual beam is in multi-mode thermal state, and the measured photon statistics can be expressed as $P(n)\approx\frac{\overline{n}}{n!}\exp (-\overline{n})$ (Possion distribution). In the former case, the value of $g^{(2)}$ equals 2; while in the latter, in the sense of realizing quantum interference among multiple sources, the temporal coherence degree is low, and the value of $g^{(2)}$ approaches 1~\cite{Mandel}.

To obtain individual beam with high temporal coherence, pulsed laser are often chosen to serve as the pump source, because photon pairs from a continuous wave laser pumped SPE always have perfect spectral correlation.
For the pulse pumped SPE, generating spectral factorable two-photon state by using narrow band optical filters and by engineering the dispersion property of nonlinear media has been extensively studied~\cite{Rarity98,ou99,Grice01,shapiro02,Uren05,Palmett07,Mosley08,li08a-opex,cohen09}. In this process, the pulsed optical fields, including the mode-locked pump pulse train and individual signal or idler fields, will acquire certain amount of chirp during propagation in a transparent medium due to the effect of chromatic dispersion and Kerr nonlinearity, particularly for the ultrashort femto-second pulses. However, for the related work presented so far, the pump is treated as a transform limited pulse, the influence of the chirp of pulsed pump has not been investigated. Additionally, the impact of the chirp of pulsed individual signal or idler beam on the temporal coherence property has yet to be studied as well~\cite{Shih02}.

In this paper, to the best of our knowledge, we study the effect of chromatic dispersion induced chirp on the temporal coherence property of individual beam generated from a SPE process for the first time. Based on SFWM in dispersion shifted fiber (DSF), two series of experiments are investigated by introducing different amount of chirp into either the pulsed pump or individual signal (idler) beam. In the first one, $g^{(2)}$ of the
filtered individual signal beam is measured as a function of the chirp of pump. The results demonstrate that the chirp of pump pulses decreases the degree of temporal coherence. In
the second one, a Hong-Ou-Mandel (HOM) type two-photon interference experiment with the signal
beams generated in two different fibers are analyzed and conducted. The results illustrate that the chirp of individual beam does not change the temporal coherence degree, but affect the mode matching. To achieve high visibility, in addition to improving the coherence degree by minimizing the chirp of pump, mode matching should be optimized by properly managing the chirps of individual beams. Our investigations are useful for quantum state engineering and quantum information processing.

The rest of the paper is organized as follows. In Sec. II, we briefly introduce basic principle of the experiment. In Sec. III, using pulse pumped SFWM in 300 m DSF, we study the temporal coherence property by measuring $g^{(2)}$ of filtered individual signal (idler) beam as the chirp of pump is varied. When the chirp of pump pulses is taken into account, we find that the two-photon joint spectral intensity (JSI) function $\left| F(\omega_i ,\omega _s)\right| ^2$ is not suitable for precisely characterizing the factorability of photon pairs, because the information of chirp contained in joint spectral amplitude (JSA) function $F(\omega_i ,\omega _s)$ vanishes in JSI. However, the measured $g^{(2)}$ of individual beam, depending on JSA and the spectrum of individual beam shaped by a filter, is a real reflection of the factorability of detected photon pairs. To obtain high temporal coherence, the chirp of pump should be minimized. In Sec. IV, using signal beams respectively generated from two 300m DSFs, we investigate a HOM type two-photon interference experiment by varying the chirps of individual beams to better understand the influence of the chirp on quantum interference among multiple sources. The results show that the observed interference dip is not only related to the temporal coherence degree, but also dependent on temporal mode matching determined by the chirps of individual beams. When the chirps of individual beams are not properly managed, the visibility is decreased, while the measured width of interference dip is increased. In Sec. V, we extend our investigation to a general pulse pumped spontaneous parametric process, and discuss the relation between $g^{(2)}$ and chirp. Finally, we give a brief conclusion.

\section{Basic principle of the experiment}

To obtain individual beam of photon pairs with high temporal coherence, the following two methods are often used. For photon pairs with high spectral correlation, one need to apply a filter with narrow bandwidth to signal or idler beam so that the detected photon pairs are spectrally factorable~\cite{ou99,yang}. In this case, the filtered individual beam is in single temporal mode. While for the photon pairs, which are directly in a spectral factorable state~\cite{Uren05,Mosley08}, individual signal or idler beam is in single mode, even without filtering.

In our experiment, photon pairs via SFWM are generated in DSF by a pulsed pump. The phase matching bandwidth of SFWM is very broad, signal and idler photon pairs with a relatively small detuning exhibit a strong spectral correlation~\cite{li08a-opex}. Therefore, to obtain individual beam with high coherence, a filter with narrow bandwidth should be applied in signal or idler field~\cite{li08ol}.

In order to study the influence of the chirp, we introduce the linear chirp by propagating the pulsed pump or individual signal (idler) beam through a piece of standard single mode fiber (SMF). For a given Gaussian shaped spectrum, the unchirped pulse can be expressed as $E(t)\propto \exp {(-\frac{t^2}{2T_0^2})}$, where $T_0$ associated with the spectral width $\sigma$ of optical field corresponds to the minimum pulse duration. If the Kerr nonlinearity is insignificant, after propagating through the SMF, the field evolves into $E(t,z)\propto \exp{ [-\frac{(t-\frac z{v_g})^2}{2T_0^2(1+C^2)}-i\frac{C(t-\frac z{v_g})^2}{2T_0^2(1+C^2)}]}$, where $v_g$ is the group velocity of optical field in SMF, $C=\beta _2z/T_0^2=\beta _2z\sigma^2$ is the chirp parameter, $\beta_2$ denotes second order dispersion coefficient, and $z$ is the length of the SMF~\cite{Agrawal}. As a consequence,
the pulse duration of the linearly chirped optical field, $\Delta T= \sqrt{1+C^2} T_0$, is enlarged.

\section{Temporal coherence of individual signal beam influenced by chirp of pulsed fields}

Our experimental setup is shown in Fig. 1. The signal and idler photon pairs are produced by SFWM in DSF, and the pump source is a mode-locked pulse train with a certain amount of linear chirp. In SFWM process, two pump photons at frequency $\omega _{p}$ scatter through the Kerr ($\chi ^{(3)}$)
nonlinearity of the fiber to create
energy-time entangled signal and idler photons at frequencies $\omega _s$ and $\omega _i$, respectively, such that $2\omega _{p}=\omega _s+\omega
_i$. To reliably detect the signal and idler photons, one must effectively suppress the pump photons from
reaching the detector, so the output of DSF
propagates through a dual band filter F2, which is realized by cascading wavelength-division multiplexing (WDM) filters with one channel of array waveguide gratings (AWG). The pump-rejection ratio provided by F2 is in excess of 120 dB.

The phase mismatch term of the SFWM is~\cite{li08a-opex}:
\begin{eqnarray}\label{delta-k}
\Delta K\approx 2 \gamma P_p+\frac{\beta_2}%
4\Delta ^2+\frac{\beta_2}%
2\Delta(\Omega _s-\Omega _i )+\frac{\beta_3}%
8\Delta ^2(\Omega _s+\Omega _i),
\end{eqnarray}
where $P_p$ is the peak pump power, $\gamma$ and $\beta_3$ are the nonlinear coefficient and third order dispersion coefficient of the fiber, respectively, $\Delta =\omega _{s0}-\omega _{i0}$ is the central frequency difference between signal and idler fields, $\Omega _s$ and $\Omega _i$ are related to $\omega _{s}$ and $\omega _{i}$ by $\Omega _s=\omega _s-\omega _{s0}$ and $\Omega _i=\omega _i-\omega _{i0}$. Here $\beta_2$ and $\beta_3$ are respectively associated with the dispersion slope of DSF through $\beta_2=-\frac{\lambda _{p0}^2}{2\pi c}D_{slope}(\lambda _{p0}-\lambda _0)$ and $\beta_3=(\frac{\lambda _{p0}^2}{2\pi c})^2D_{slope}$.
When the central wavelength of pump $\lambda _{p0}$ is longer than the zero dispersion wavelength $\lambda
_0$ of DSF, we get
$\beta_{2}<0$ which, in addition to a small amount of self-phase modulation and third order dispersion results in the satisfaction of $\Delta K\approx0$.

In the experiment, 300 m DSF with
 $\lambda
_0=1538\pm 2$ nm and $D_{slope}=0.075 $ ps/(nm$^2\cdot$km) is submerged in liquid nitrogen to suppress Raman scattering~\cite{li08ol}. The central wavelength and repetition rate of the mode-locked pump laser are about 1538.9 nm and 41 MHz, respectively. The fiber
polarization controller (FPC1) and a polarization beam splitter
(PBS1) are used to ensure the polarization and power adjustment of pump. Signal and idler photons co-polarized with the pump are selected by adjusting FPC2 placed in front of PBS2.
The central wavelength of F2 in signal and idler bands are 1546.9\,nm and 1530.9\,nm, respectively. Under this condition, $\Delta K\approx0$ is satisfied, and an efficient SFWM with broad band phase matching is realized. To obtain signal or idler photons with high temporal coherence, the full width at half maximum (FWHM) of the narrow band filter F2 is set to be 0.4 nm in both signal and idler bands.

To achieve the pulsed pump with different amount of linear chirp, we first obtain the initial pump pulses by passing the output of a mode-locked femto-second fiber laser through filter F1, whose central wavelength and FWHM are 1538.9 and 1 nm, respectively. Next, we introduce chirp by propagating the initial pump along a piece of standard SMF with $\beta_2$ of about $-20$ ps$^2$/km. Before launching into the SMF, the initial pump is attenuated to about 10 $\mu$W to ensure the chirp is mainly originated from the chromatic dispersion, but not Kerr nonlinearity. The quantity of the imported linear chirp can be changed by varying the length of SMF. To obtain the required power, the pump
pulses are then amplified by an erbium-doped fiber amplifier (EDFA), and cleaned up by passing
through another F1. For the SMF with length of 400 m, 600 m, 800 m, 1km, 1.2 km, and 1.4 km, respectively, we measure the pulse duration of the chirped pulses passed through EDFA and F1 by using an auto-correlator.  According to the relation between the pulse duration and chirp, $\Delta T_p=0.44\lambda _{p0}^2\sqrt{1+C_p^2}/(2\sqrt{\ln {2}}c\Delta \lambda_p)$, with $\Delta \lambda_p$ denoting the FWHM of pump, the amount of chirp of pump field $C_p$ can be deduced. Figure 2(a) plots $\sqrt{1+C_p^2}$ as a function of the length of SMFs.  Note that the pulsed pump is not transform limited even if the SMF is by passed, because quantity of chirp $C_p$ presented in the initial pump is greater than $0$.

The signal (idler) photons propagated through F2 are then counted by SPDs (id200) operated in the gated Geiger mode. The $2.5$\,ns gate pulses arrive at a rate of about $2.58$\,MHz, which is $1/16$ of the repetition rate of the pump pulses, and the dead time of the gate is set to be 10 $\mu$s. The timing of gate pulses are adjusted by a digital delay generator to coincide with the arrival of signal (idler) photons. To measure the intensity correlation function $g^{(2)}$, photons in signal (idler) band propagate through a 50-50 beam splitter (BS) and then are detected by SPD1 and SPD2, respectively. In the measurement, path matching is required to ensure the photons detected by two SPDs are produced by the same pump pulse.
Because the coherence property of filtered individual signal and idler photons is identical in our experiment, in the following description, we only present the results of signal photons, and refer to the signal photons as signal beam.

Before presenting experimental data, let's briefly analyze the the dependence of measured $g^{(2)}$ of signal beam by taking the chirps of pulsed pump and individual beams into account. The strong pump pulses with a linear chirp $C_p$ and Gaussian shaped spectrum remain classical, and can be written as
\begin{eqnarray}
E_{p}^ + = E_{p0} e^{ - i\gamma P_p z} \int {d\omega _{p} e^{ - \frac{(\omega _{p} -
\omega _{p0} )^2}{2\sigma _p^2
} (1+iC_p)} e^{ik_{p} z - i\omega _{p} t}},
\end{eqnarray}
where $P_{p} \propto \sigma _{p}^{2}E_{p0}^{2}/\sqrt{1+C_p^2}$, and $\sigma _{p}$ denotes the bandwidth of pump. In the Heisenberg picture, and in the low gain regime, the field operator of co-polarized signal beam at the output of DSF is ~\cite{yang}
\begin{eqnarray}
a(\omega _{s})={a_{0}(\omega _{s})}+\frac{G}{
\sigma _{p}} \int {d\omega_{i} F (\omega_{i},\omega _{s} ) a_{0}^{\dag
}(\omega_{i} )}+o(G),
\end{eqnarray}
where $a_{0}(\omega _{s})$ and $a_{0}^{\dag
}(\omega_{i} )$ are annihilation and creation operators of the vacuum fields at $\omega_s$ and $\omega_i$, respectively, $G\propto \gamma P_p L \sqrt{1-iC_p}$ is proportional to the gain of SFWM. The JSA has the form: $F (\omega_{i},\omega _{s} ) \propto \exp [ - \frac{1+iC_p}{{4\sigma _p^2 }}(\Omega_i
+ \Omega _{s} )^2] \mathrm{sinc} ( \frac{1}{{2 }}\Delta K L )$,
where $L$ is the length of DSF.
Since the value of the term, $\beta_2\Delta ^2 /4+2\gamma P_p$, in Eq. (\ref{delta-k}) is negligibly small, using the approximation $\mathrm{sinc} ( \Delta K L/2 )\approx\exp [ { - \frac{1}{{2 }}( \frac{{\Omega_i  }}{A} + \frac{\Omega _s }{B} )^2 } ]$, we obtain the expression of JSA~\cite{li08a-opex}
\begin{eqnarray}\label{jointspectra}
F(\omega_{i} ,\omega _{s})
 \propto  \exp [ - \frac{1+iC_p}{{4\sigma _p^2 }}(\Omega_i
+ \Omega _{s} )^2]\exp [ { - \frac{1}{{2 }}( \frac{{\Omega_i  }}{A} + \frac{\Omega _s }{B} )^2 } ],
\end{eqnarray}
where the coefficients
$ A = 6.44 /[L( \beta_3 \Delta ^2 /4 - \beta_2 \Delta )]\gg \sigma _p$ and
$ B = 6.44/ [L(\beta_3 \Delta ^2 /4 + \beta_2 \Delta )]\gg \sigma _p$, are proportional to the bandwidth of phase matching function in idler and signal bands.
Eq. (\ref{jointspectra}) shows that the directly generated signal and idler photon pairs are spectrally correlated, and the correlation increases with the increase of $C_p$. However, the chirp induced correlation vanishes in the expression of JSI function
\begin{eqnarray}\label{jsi}
\left| F(\omega_i ,\omega _s)\right| ^2
 \propto  \exp [ - \frac{(\Omega_i
+ \Omega _{s} )^2}{{2\sigma _p^2 }}]\exp [ { - ( \frac{{\Omega_i  }}{A} + \frac{\Omega _s }{B} )^2 } ].
\end{eqnarray}
Therefore, even though the feature of JSI can be experimentally observed by a coincidence measurement with two-dimension scanning in the signal and idler bands, and was often used to characterize the factorability of JSA~\cite{Mosley08,li08a-opex}, our analysis shows that JSI is not suitable for accurately characterizing the spectral correlation property of photon pairs for the case of $C_p\neq0$.

After passing through the filter F2, the field operator of the signal beam can be written as
\begin{equation}\label{sig-operator}
   A_{s}^+(t)=\frac{1}{{\sqrt {2\pi } }}\int {d\omega _s f (\omega _s )a(\omega _s )e^{ - i\omega _s t} },
\end{equation}
where
$ f(\omega _{s})=\exp{[-\Omega _{s} ^2(1+iC_s^{\prime})/(2\sigma_{s}^{2})]} $ describes spectrum of F2 in signal band and the chirp of signal beam $C_s^{\prime}$ introduced by transmission fibers.
In our experiments, since the bandwidth of filtered signal beam $\sigma_{s}$ is much smaller than $A$ and $B$, for the expression of  $A_{s}^+(t)$ in Eq. (\ref{sig-operator}), we can apply the approximation $\exp [ { - \frac{1}{{2 }}( \frac{{\Omega_i  }}{A} + \frac{\Omega _s }{B} )^2 } ]\approx 1$ (see Eq. (\ref{jointspectra})) to simplify the analysis.

In the photon counting measurement, the filed operators of signal beam detected by SPD1 and SPD2 can be written as
\begin{eqnarray}\label{}
 E_{1}^ {+}  (t) &=& \sqrt{T}\sqrt{\eta_{1}} A_{s}^+(t),\cr
 E_{2}^ {+}  (t) &=& i\sqrt{R}\sqrt{\eta_{2}} A_{s}^+(t),
\end{eqnarray}
where $R$ and $T$ denote the reflectivity and transmissivity of the BS,
$\eta _j$ (j=1,2) denotes the total detection efficiency determined by the efficiencies of F2 and SPDs. Accordingly, the coincidence rate between SPD1 and SPD2 is
\begin{eqnarray}\label{N12}
 N_{12}= \int {dt_1 dt_2 \left| {\left| {E_1^ +  (t_1 )E_2^ +  (t_2 )\left| 0 \right\rangle } \right|} \right|^2 },
\end{eqnarray}
and the intensity correlation function can be calculated and expressed as
\begin{equation}\label{g2}
   g^{(2)} = \frac{N_{12}} {N_1 N_2} = 1+\frac{\int {d\omega _s d\omega _s '\left| {\int {d\omega _i f(\omega _s )F^* (\omega _i ,\omega _s )f(\omega _s ')F(\omega _i ,\omega _s ')} } \right|^2 }}{\left|\int {d\omega _s }d\omega _i \left| {f(\omega _s )F(\omega _i ,\omega _s )} \right|^2 \right|^2}
   = 1+ \frac{{1}}{{\sqrt {1 + \frac{{\sigma _{s}^2 }}{{2\sigma _p^2 }}\left( {1 + C_p^2 } \right)} }},
\end{equation}
where $N_1=\int_{-\infty }^{+\infty } {dt\left\langle 0 \right|E_1^- (t)E_1^ +  (t)\left| 0 \right\rangle }$ and $N_2=\int_{-\infty }^{+\infty } {dt\left\langle 0 \right|E_2^- (t)E_2^ +  (t)\left| 0 \right\rangle }$ are the single count rates of SPD1 and SPD2, respectively. Notice $g^{(2)}\leq 2$ because of the Schwatz inequality, the equality holds if and only if JSA $ F(\omega_i ,\omega _s)$ can be factorized. For the given DSF, $F(\omega_i ,\omega _s)$ is not factorable, however, we can make the product of $f(\omega _s )$ and $F(\omega_i ,\omega _s)$ approximatively factorable by applying the narrow band filter in signal field. Moreover,
it is clear that $g^{(2)}$ of individual signal beam depends on the chirp of pump $C_p$, but has no relevance to the chirp of individual beam $C_s^{\prime}$. When $C_p$ increases, the factorability of the product $f(\omega _s ) F(\omega_i ,\omega _s)$ decreases, unless the bandwidth of filtered signal beam $\sigma _{s}$ is further reduced.

To verify the above analysis, we experimentally measure $g^{(2)}$ of the filtered signal beam as a function of the chirp of pump. In the measurement, the peak power of
the pump is fixed at about $1$\,W to ensure the phase matching described by Eq. (\ref{delta-k}) does not change, and the chirp parameter $C_p$ is varied by propagating the initial pulses through
SMF with different lengths. As shown in Fig. 2(b), one sees that
$g^{(2)}$
decreases with the increase of $\sqrt{1+C_p^2}$. Additionally, we compute $g^{(2)}$ by
substituting the experimental parameters into Eq. (\ref{g2}). It is clear that the calculations agree with the experimental results.

\section{HOM type two-photon interference influenced by the chirp of individual beams}

Having investigated the dependence of the temporal coherence of individual signal beam,
we study a HOM-type two-photon interference formed by signal photons generated from independent sources to better understand the influence of chirp. The experimental setup is shown in Fig. 3. The filtered signal$_1$ and signal$_2$ beams with identical spectrum are obtained by passing the pulsed pump through a 50-50 BS (BS1) and then
pumping DSF1 (300\,m) and DSF2 (300\,m), respectively. After propagating through transmission fibers, the two signal beams are carefully path matched and
simultaneously fed into a 50-50 BS (BS2) from two input ports, respectively. Before coupling into BS2, signal$_2$ beam originated from DSF2 is
delayed by the reflector mirrors mounted on a translation stage. The two output ports of BS2 are detected by SPD1 and SPD2, respectively. Since we have previously demonstrated the dependence of observed visibility of interference upon $g^{(2)}$ of individual beam in Ref.\cite{li08ol}, here we will focus on studying the influence of the chirps of individual signal beams. Therefore, the chirp of pump is fixed, while different amount of chirps are introduced to signal beams by passing through transmission SMFs with different dispersion properties and with different lengths.

Let's analyze the experiment at first. The field operators of chirped signal$_1$ and signal$_2$ beams can be expressed as $A_{s1} ^+(t)$ and $A_{s2} ^+(t)$ (see Eq.(\ref{sig-operator})), respectively.
The function describing filters and transmission media of $A_{s1}^+(t)$ and $A_{s2}^+(t)$ are
\begin{eqnarray}\label{}
f_1(\omega _s ) = \exp \{  - \frac{{\Omega _s^2 (1 + iC_{s1} ')}}{{2\sigma _s^2 }}\}
\end{eqnarray}
and
\begin{eqnarray}\label{}
f_2(\omega _s ) = \exp \{  - \frac{{\Omega _s^2 (1 + iC_{s2} ')}}{{2\sigma _s^2 }}\},
\end{eqnarray}
respectively, where $C_{s1} '$ and $C_{s2} '$ stand for the chirps introduced to the two signal beams.
When the two signal fields are combined at BS2, the relative delay between them is $\delta\tau=2\delta l/c$, where $\delta l$ is the difference
in readings of the translation stage.

In the photon counting measurement, the fields operator detected by SPD1 and SPD2 are:
\begin{eqnarray}\label{}
    E_{1^{\prime} } ^{+ } (t) &  =&  \sqrt {\eta _{1^{\prime} }}\left[ {\eta \sqrt T A_{s1} ^+(t+\delta\tau) + i\sqrt R A_{s2}^+ (t) } \right],\cr
    E_{2^{\prime} }
   ^{ + } (t) &  =&  \sqrt {\eta _{2^{\prime} } } \left[ {i\eta \sqrt R A_{s1}^+ (t+\delta\tau)  + \sqrt T A_{s2}^+ (t) } \right],
\end{eqnarray}
where $\eta$ denotes the ratio between the intensities of signal$_1$ and signal$_2$ beams at the input ports of BS2, and $\eta _{{1^{\prime}(2') }}$ is total detection efficiencies of SDP1 (SPD2). Accordingly,
the coincidence rate of SPD1 and SPD2 is written as
\begin{eqnarray}\label{HOM}
N_{1'2'}(\delta \tau) &=& {\frac{{N_{1'} N_{2'}TR (1 + \eta ^4 )(g^{(2)}  - 1 )}}
{{\left( {\eta ^2 T + R} \right)\left( {\eta ^2 R + T} \right)}} \left[ {1 - S\frac{{2\eta ^2 \xi (\delta \tau)}}{{(1 + \eta ^4 )}}} \right]} + N_{1'} N_{2'}
\end{eqnarray}
with
\begin{equation}\label{dipwidth}
 \xi (\delta \tau )=\exp \left\{ { - \frac{{\delta \tau ^2 \sigma _s^2(g^{(2)}  - 1 ) ^2S^2}}{{2}}} \right\},
\end{equation}
where $N_{1^{\prime} }=\int_{-\infty }^{+\infty } {dt\left\langle 0 \right|E_{1'}^- (t)E_{1'}^ +  (t)\left| 0 \right\rangle }$ and $N_{2^{\prime} }=\int_{-\infty }^{+\infty } {dt\left\langle 0 \right|E_{2'}^- (t)E_{2'}^ +  (t)\left| 0 \right\rangle } $ are the single count rates of SPD1 and SPD2, respectively.
Note that the coefficient
\begin{equation}\label{S}
 S= \sqrt {\frac{\tau_s^2+\frac{1}{2}\Delta T _p^2 }{\tau_s^2+ \frac{1}{4}\tau_s^2(C_{s1} ' - C_{s2} ')^2 +\frac{1}{2}\Delta T_p^2}} \leq 1
\end{equation}
in Eqs. (\ref{HOM}) and (\ref{dipwidth}) is associated with the temporal mode matching between the chirped signal fields $A_{s1} ^+(t)$ and $A_{s2} ^+(t)$, where $\tau_s=1/\sigma_s$ is the coherence time of signal field. The equality $S=1$ only holds for $C_{s1} ' = C_{s2} '$.

From Eqs. (\ref{HOM}) and (\ref{dipwidth}), one sees that for the coefficient $S$ with a certain value, the coincidence rate $N_{1'2'}$ increases with the increase of time delay $\delta \tau^2$, and the smallest $N_{1'2'}$ will be achieved at $\delta \tau=0$. Therefore, a interference dip should be observed by measuring coincidences versus
the position of the translation stage.

To figure out more observable features of the two-photon interference dip, let's analyze its FWHM and visibility, $\Delta \tau$ and $V$, respectively. According to Eqs. (\ref{HOM}) and (\ref{dipwidth}), taking the relative delay quantities $\delta \tau=\frac{-\Delta \tau }{2}$ and $\delta \tau=\frac{\Delta \tau }{2}$, which correspond to the average of the measured maximum and minimum coincidence rate, i.e. $N_{{1^{\prime} }{2^{\prime} }} (\frac{-\Delta \tau }{2})=N_{{1^{\prime} }{2^{\prime} }} (\frac{\Delta \tau}{2})=\frac{N_{{1^{\prime} }{2^{\prime} }} (\delta \tau \rightarrow\infty )+N_{{1^{\prime} }{2^{\prime} }} (\delta \tau =0)}{2}$, we obtain the expression of the FWHM:
\begin{equation}\label{FWHM}
\Delta \tau= \frac{{2\sqrt {2\ln 2} }}{{ (g^{(2)} -1) \sigma _s S}}.
\end{equation}
Defining the visibility as $
V=\frac{N_{1'2'} (\delta \tau \rightarrow\infty )-N_{1'2'}(\delta \tau =0)}{N_{{1^{\prime} }{2^{\prime} }} (\delta \tau \rightarrow\infty )}$, we find the dependence of $V$ and $\Delta \tau$ is different. $\Delta \tau$ can be fully determined by the coefficient $S$, bandwidth and intensity correlation function of the signal beams, $\sigma _s$ and $g^{(2)}$, but $V$ also relies on the parameters $R$ ($T$) and  $\eta$, which respectively denote reflectivity (transmissivity) of BS2 and ratio between the intensity of two signal beams. For clarity, substituting the parameters $R=T$ and $\eta=1$, into Eq. (\ref{HOM}),  which correspond to the maximized visibility for $g^{(2)}$ and $S$ with certain values, we obtain the expression of visibility:
\begin{equation}\label{visibility}
V= \frac{(g^{(2)} -1) S }{{g^{(2)}  + 1}}.
\end{equation}
Note that under the condition of $S=1$, the highest visibility would be $V=33 \%$ for $g^{(2)}=2$ because of the thermal nature~\cite{ou99,li08ol}. However, if the two-photon interference can be measured by gating the idler fields of DSF1 and DSF2, the signal fields would be single photon Fock states, and the visibility theoretically quantified by $V=g^{(2)} -1$ can be expected to reach $100 \%$~\cite{ou99,Fulconis06,Mosley08,li08b-opex}.

According to Eqs. (\ref{FWHM}) and (\ref{visibility}), one sees that the interference dip with a increased FWHM and reduced visibility will be observed for the case of $S <1$ ($C_{s1} '\neq C_{s2} '$). Therefore, in order to improve the visibility of quantum interference among multiple sources, apart from making effort to obtain individual beams with high coherence, the chirps of pulsed individual beams, $C_{s1} '$ and $ C_{s2} '$, should be properly managed.
It is worth noting that the FWHM given by Eq. (\ref{FWHM}) is not a real reflection of pulse duration of signal fields due to the slow response time of SPDs~\cite{Steinberg92}.

Having analyzed the performance of the HOM interferometer, we experimentally measure the coincidences versus
position of the translation stage in three cases. In the first case, for both the signal$_1$ and signal$_2$ beams, the lengths of their single mode transmission fibers are so short that the induced chirps are neglectable, i.e. $C_{s1} '= C_{s2} '\approx0$; in the second case, both signal$_1$ and signal$_2$ beams propagate through the same kind of standarded SMF with length of about 1.5 km, i.e., $C_{s1} '\approx C_{s2} '\approx-1.2$; while in the third case, signal$_1$ and signal$_2$ transmit through 1.4 km standard SMF and 280 m dispersion compensation fiber ($\beta_2\approx 100$ ps$^2$/km), respectively, i.e., $C_{s1} '\approx -1.1$, $C_{s2} '\approx 1$. For the first and second cases, we have $S=1$, while for the last case, we have  $S\approx 0.7$. During the measurement, the average power of the pump with the currently achievable minimum chirp is about 0.6 mW, and the measured $g^{(2)}$ is about 1.94 for both signal beams, which is irrelevant to the chirp of individual beam. Moreover, to observe the maximized visibility, the intensities of signal$_1$ and signal$_2$ beams are adjusted to be equal ($\eta=1$), and the polarization of signal$_1$ is adjusted by FPC to ensure the polarizations of the two signal fields involved in interference are the same. Furthermore, to conveniently compare the data obtained in different cases, the directly measured coincidences $N_{{1^{\prime} }{2^{\prime} }}$ are normalized by the calculated accidental coincidences, which is the product of single count rates of two SPDs, $N_{1^{\prime} }N_{2^{\prime} }$.

Figure 4 shows the experimental results. One sees that the biggest values of normalized coincidence $N_{{1^{\prime} }{2^{\prime} }}/(N_{1^{\prime}}N_{2^{\prime}})$ are the same in three cases. For $S=1$, one sees that the data obtained under the conditions of $C_{s1} '= C_{s2} '\approx0$ and $C_{s1} '\approx C_{s2} '\approx-1.2$, respectively, is almost overlapped, the observed visibility and FWHM of the dip, $V$ and $\Delta \tau$, are about $ 31 \%$ and 13 ps, respectively. However, for $S\approx 0.7$, $V$ is reduced to about $ 18 \%$, and $\Delta \tau$ is increased to about 17 ps. Moreover, we compute the value of the normalized coincidence by substituting the experimental parameters into Eq. (\ref{HOM}), and the calculations agree with the measured results quite well. The results demonstrate that for individual beams with high temporal coherence, the high visibility can be preserved by properly managing the chirps of individual beams, otherwise the visibility and FWHM of interference dip will be decreased and increased, respectively.

\section{Discussion}
Finally, we would like to mention that our study can be extended
to other pulse pumped SPEs. Using the general expression of JSA for SPDC and SFWM (see Ref.\cite{shapiro02} and Eq. (\ref{jointspectra})), the intensity correlation function of individual signal or idler beam can be respectively expressed as
\begin{equation}\label{general-bunching-s}
 g^{(2)}_s  =1+ \frac{{\sqrt {1 + \frac{{a\sigma _p^2 }}{{A'^2 }} + \sigma _s^2 \left( {\frac{1}{A'} - \frac{1}{B'}} \right)^2 } }}{{\sqrt {1 + \frac{{a\sigma _p^2 }}{{A'^2 }} + \frac{{\sigma _s^2 }}{{A'^2 }} + \frac{{\sigma _s^2 }}{{B'^2 }}\left( {1 + \frac{{a\sigma _p^2 }}{{A'^2 }}} \right) + \frac{{\sigma _s^2 }}{{a\sigma _p^2 }}\left( {1 + C_p^2 } \right)} }}
\end{equation}
and
\begin{equation}\label{general-bunching-i}
 g^{(2)}_i  =1+ \frac{{\sqrt {1 + \frac{{a\sigma _p^2 }}{{B'^2 }} + \sigma _i^2 \left( {\frac{1}{B'} - \frac{1}{A'}} \right)^2 } }}{{\sqrt {1 + \frac{{a\sigma _p^2 }}{{B'^2 }} + \frac{{\sigma _i^2 }}{{B'^2 }} + \frac{{\sigma _i^2 }}{{A'^2 }}\left( {1 + \frac{{a\sigma _p^2 }}{{B'^2 }}} \right) + \frac{{\sigma _i^2 }}{{a\sigma _p^2 }}\left( {1 + C_p^2 } \right)} }},
\end{equation}
where $\sigma _i$ denotes the bandwidth of filter in idler band; $A'$ and $B'$ describe the bandwidth of phase matching function in signal and idler bands, respectively; $a$ is equal to $1$ and $2$ for SPDC and SFWM, respectively. Equations (\ref{general-bunching-s}) and (\ref{general-bunching-i}) indicate that despite properly designing the dispersion property of nonlinear media to regulate the ratio of $A'/B'$~\cite{Uren05,Palmett07}, the temporal coherence degree of individual beam can be changed by manipulating the chirp parameter $C_p$. From this point of view, the chirp of pump might be viewed as another degree of freedom for engineering the quantum state of photons from spontaneous parametric processes.

Since the intensity correlation function of individual beam is associated with the spectral correlation of photon pairs (see Eq. (\ref{g2}), according to Eqs. (\ref{general-bunching-s}) and (\ref{general-bunching-i}), one sees that the present of $C_p$ is good for increasing the spectral correlation of photon pairs. However, it seems that $C_p$ is adverse for directly generating photon pairs in a spectral factorable state, which is in principle filter free ($\sigma_p \ll \sigma_s$ and $\sigma_p \ll \sigma_i$)~\cite{Uren05,Palmett07}. With the increase of the chirp $C_p$, to maintain the degree of temporal coherence, the filter with a narrower bandwidth should be required. Therefore, to improve the coherence degree of individual beam, it is necessary to decrease the spectral correlation of photon pairs by reducing the chirp quantity of pump pulses.

\section{Conclusion}
In conclusion, by introducing different amount of chromatic dispersion induced chirp into either the pulsed pump or signal (idler) beam, we have investigated the effect of chirp on the temporal coherence property of individual beam generated from a spontaneous parametric process for the first time. Using the pulse pumped SFWM in 300 m DSF, we have studied the degree of temporal coherence as a function of the chirp of pump. The degree of temporal coherence is characterized by the intensity correlation function of individual beam $g^{(2)}$, which is a real reflection of the factorability of detected photon pairs, and the experimental results demonstrate that the chirp of pump pulses decreases the degree of temporal coherence. To improve the temporal coherence degree, the chirp of pump should be minimized. Moreover, to better understand the influence of the chirp on the quantum interference among multiple sources, a HOM type two-photon interference experiment with the signal beams generated in two different fibers is analyzed and conducted. The results illustrate that the chirp of individual beams does not change the temporal coherence degree, but affect the temporal mode matching. If the chirp of individual beams is not properly managed, the visibility will be decreased. To achieve high visibility, apart from improving the coherence degree, the mode matching should be optimized by matching the chirps of individual beams. We believe our study can be extended
to other pulse pumped spontaneous parametric emission processes, therefore, is very useful for quantum state engineering and for quantum information processing.

\begin{acknowledgments}

We thank Professor Z. Y. Ou and Professor Ching-Yue Wang for fruitful discussions. This work was supported in part by the NSF of China
(Grant No. 11074186), 111 Project (Grant No. B07014), and the State Key
Development Program for Basic Research of China (Grant No. 2010CB923101)

\end{acknowledgments}



\newpage


\begin{figure}
\includegraphics[width=8cm]{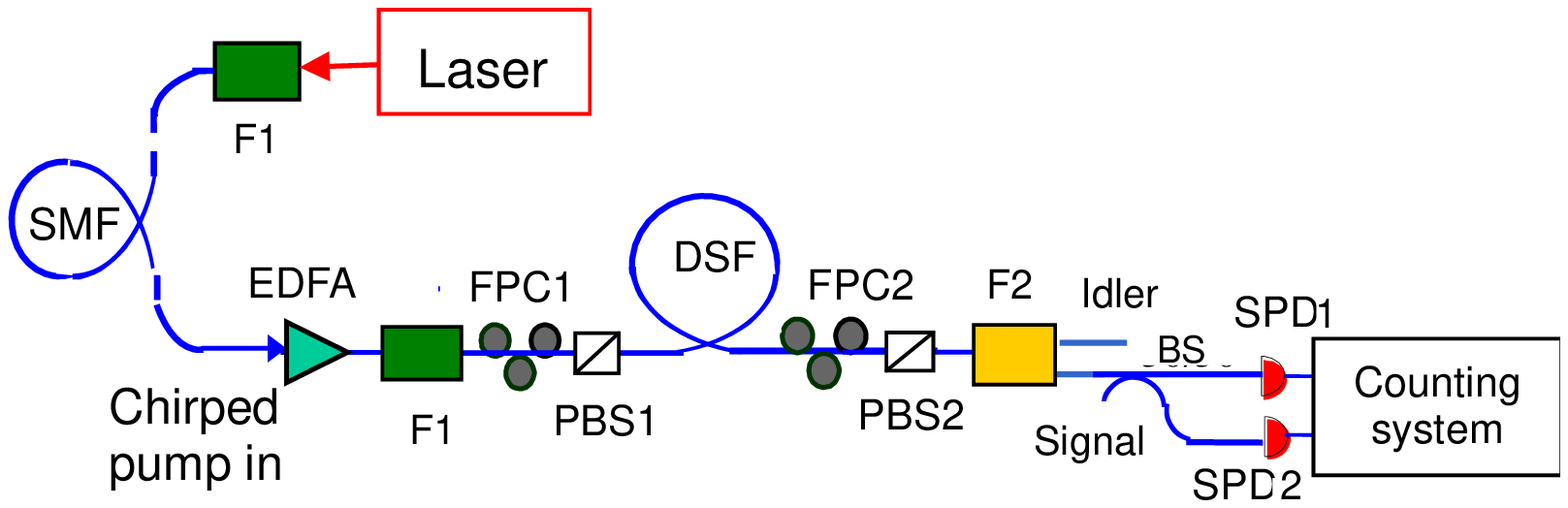}
\caption{(Color online) Experimental setup. SMF, single mode fiber; FPC, fiber
polarization controller; PBS, polarization beam splitter; F,
filter; BS, beam splitter. }
\label{}
\end{figure}

\begin{figure}
\includegraphics[width=8cm]{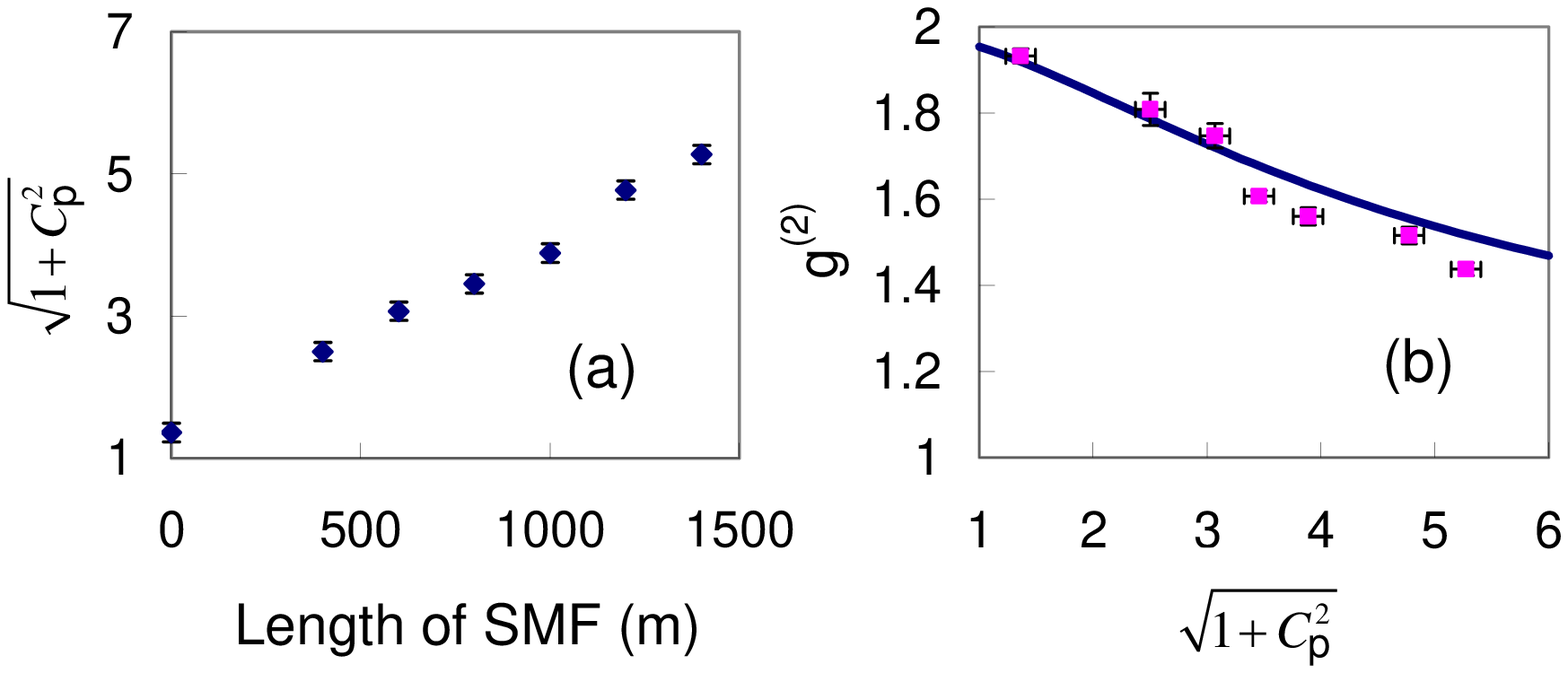}
\caption{(Color online) (a) The chirp quantity of pump versus the length of SMF. (b) The intensity correlation function $g^{(2)}$ of filtered individual signal beam
as a function of the chirp of pump. The solid curve is the calculated result by substituting experimental parameters into Eq. (\ref{g2}). }
\label{}
\end{figure}

\begin{figure}[htb]
\includegraphics[width=8cm]{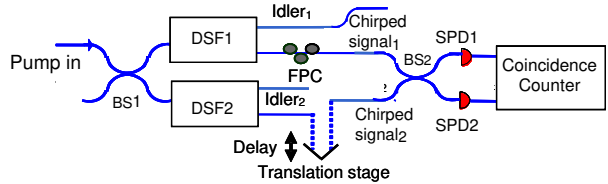}
\caption{(Color online) Experimental setup of two-photon Hong-Ou-Mandel (HOM)
interferometer. }
\label{}
\end{figure}

\begin{figure}[htb]
\includegraphics[width=8cm]{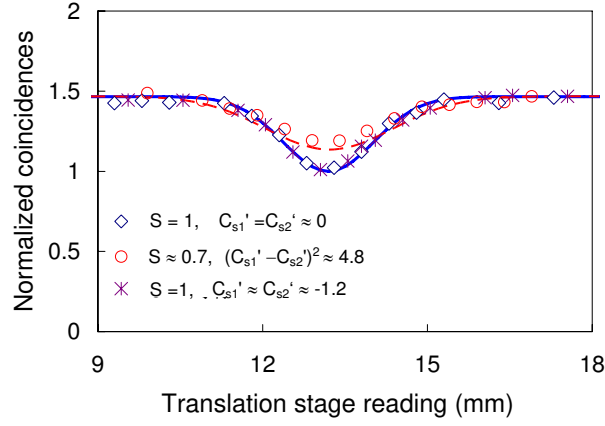}
\caption{(Color online) The normalized two-fold coincidences $N_{{1^{\prime} }{2^{\prime} }}/(N_{1^{\prime}}N_{2^{\prime}})$ versus
position of the translation stage in three cases. Error bars are about the same as the size of data points. The solid and dashed curves for the case of $S=1$ and $S\approx0.7$ are obtained by substituting the measured result of $g^{(2)}=1.94$ and other experimental parameters into Eq. (\ref{HOM}), no fitting parameters are used. }
\label{}
\end{figure}

\end{document}